# An updated comparison of model ensemble and observed temperature trends in the tropical troposphere

## Stephen McIntyre[(1)], Ross McKitrick [(2)]

(1): Climate Audit, stephen.mcintyre@utoronto.ca; (2) Univsersity of Guelph

## Abstract

A debate exists over whether tropical troposphere temperature trends in climate models are inconsistent with observations (Karl et al. 2006, IPCC (2007), Douglass et al 2007, Santer et al 2008). Most recently, Santer et al (2008, herein S08) asserted that the Douglass et al statistical methodology was flawed and that a correct methodology showed there is no statistically significant difference between the model ensemble mean trend and either RSS or UAH satellite observations. However this result was based on data ending in 1999. Using data up to the end of 2007 (as available to S08) or to the end of 2008 and applying exactly the same methodology as S08 results in a statistically significant difference between the ensemble mean trend and UAH observations and approaching statistical significance for the RSS T2 data. The claim by S08 to have achieved a "partial resolution" of the discrepancy between observations and the model ensemble mean trend is unwarranted.

## Background

The controversy over claims of inconsistency between tropical troposphere temperature trends in climate models and observations was discussed by the U.S. Climate Change Science Program ("CCSP") (Karl et al 2006) and IPCC (2007), as well as by Douglass et al 2007 and S08. We will not try to summarize it here. Although CCSP had an extensive commentary on whether models and satellite observations were "consistent" and its Statistical Appendix to Chapter 5 recommended methods for comparing trends, they did not undertake a test of whether the difference between the model ensemble mean trend and observations was significant. Douglass et al 2007 proposed such a test and concluded that the difference was significant.

S08 criticized the Douglass et al methodology, proposing instead a modified *t*-test (denoted $d_1^*$) defined in their equation (12) as:

$$d_1^* = \frac{<<b_m>> - b_0}{\sqrt{\frac{1}{n_m} s(<b_m>)^2 + s(b_0)^2}} \qquad (1)$$

where $b_0$ is the observed trend, $s(b_0)^2$ is the estimated variance of $b_0$, $<<b_m>>$ is the average trend over all models, $s(<b_m>)^2$ is the square of the 'inter-model standard deviation of ensemble-mean trends' and $n_m$ is the number of models (19 in S08). S08 Table 1 reports ensemble mean trends $<<b_m>>$ and standard deviation estimates $s(<b_m>)$. For the T2LT level, the model-average trend $<<b_m>>$ is reported as

0.215 deg C/decade and the inter-model standard deviation $s(<b_m>)$ is 0.092 deg C/decade (with corresponding values for the T2 level.)

Douglas et al. had used the same *form* of test, but had treated the observed trend $b_0$ as non-stochastic. This yielded the following t-statistic for the difference between the ensemble average trend and observed trend (expressed here in S08 notation):

$$d_1 = \frac{<<b_m>> - b_0}{s(b_m)/\sqrt{n_m}} \qquad (2)$$

which is just (1) with $s(b_0) = 0$.

We agree with S08 that the observed trend $b_0$ should be treated as stochastic. We also agree with their observation that the autocorrelation of trend residuals should be considered in estimating the standard error of the observed trend. This is where the differences between Douglass et al and S08 results arise, not from a difference between them in how the variance of the model ensemble mean trend was estimated.

A simple ordinary least squares (OLS) trend through the UAH T2LT series in the S08 reporting period of 1979-1999 (using data downloaded in January 2009) yields a trend coefficient of 0.059 deg C/decade and a trend standard error of 0.031 deg C/decade. As noted in S08, the trend residuals are highly autocorrelated in each of the observation data sets. For the UAH T2LT series they reported an AR1 coefficient of $r_1 = 0.891$ (S08, Table I line 9), with similarly high coefficients for other observational data sets. Failure to account for autocorrelated trend residuals can lead to substantially understating the standard error of the trend estimate. Various procedures have been proposed over the years for treating this problem. S08 used a method recently discussed in CCSP chapter 5 Appendix, which cited prior discussion by S08 coauthors (Santer et al. 2000), though the form of adjustment has a long prior history, dating back to at least Quenouille (1952). Under the assumption that autocorrelation is strictly first-order, an "effective degrees of freedom" $n_e$ is calculated according to:

$$n_e = n_t \frac{1-r_1}{1+r_1} \qquad (3)$$

where $n_t$ is the number of degrees of freedom in the simple OLS calculation. For the UAH T2LT data set, this adjustment reduced the effective number of degrees of freedom from 250 (=252-2) to only 14.5 (Table 1 line 9). In their Supplementary Information, S08 cite Nychka et al (2000) as authority for this form of adjustment; however, we note that the form of adjustment in Nychka et al. differs from this.

S08 then adjusted the standard error of the observed trend by multiplying by the square root of the ratio of the unadjusted and effective degrees of freedom:

$$s_{adj}(b_0) = s_{OLS}(b_0)\sqrt{\frac{N-2}{n_e-2}}. \tag{4}$$

In the case at hand, the standard error of the OLS trend estimate is multiplied by $\sqrt{(252-2)/(14.5-2)}$ yielding an adjusted standard error of 0.138 for the UAH T2LT observed trend (see S08 Table I line 9 column 2), instead of the OLS standard error of 0.031 deg C/decade. This 4-fold increase in $s(b_0)^2$ increases the denominator in (1), reducing the $d_1^*$ statistic by enough that the difference between the ensemble mean trend and observations becomes insignificant. For instance, the UAH T2LT series $d_1^*$ score falls from 7.16 to 1.11 (see S08 Table III).

This increase in the denominator applies not just to comparisons between the observed trend and the model ensemble average trend, but to comparisons between the observed trend and zero. We will return to this point later.

S08 assessed $d_1^*$ statistical significance by comparing the $d_1^*$ values to percentiles of standard *t* distributions, reporting when a hypothesis was "rejected at the 5% level or better." CCSP (Appendix to Chapter 5) had stated that a one-sided t-test should be applied when "we expect a trend in a specific direction." Given the lengthy and occasionally acrimonious dispute over whether models over-state observations (and not whether they under-state observations), this clearly indicates the appropriateness of a one-sided t-test, thereby yielding the 95<sup>th</sup> percentile as the appropriate benchmark for "rejection at the 5% level." Nevertheless, S08 stated that they had "no expectation a priori regarding the direction of the trend difference" between models and observations and therefore applied a two-sided *t*-test (equivalent to the 97.5<sup>th</sup> percentile for 5% rejection).

Using data up to the end of 1999, S08 determined that none of the $d_1^*$ scores calculated using the expression in their equation 12 (our equation 1) exceeded the 97.5<sup>th</sup> percentile of the *t*-distribution for any of the four observational data sets (UAH/RSS; T2/T2LT), on which basis they concluded that there was no "statistically significant" difference between the ensemble average trend and any of the observation data sets.

While S08 only reported model ensemble mean trend results up to 1999, observations are available on a timely basis up to the present (S08 noted that results up to the end of 2007 were available to them). S08 rationalized the use of data ending in 1999 as follows:

> Since most of the 20CEN experiments end in 1999, our trend comparisons primarily cover the 252-month period from January 1979 to December 1999, which is the period of maximum overlap between the observed MSU data and the model simulations.

The sensitivity of S08 results to the use of updated data is an obvious question. Given the relative continuity of forcings, it is reasonable to extrapolate model trends so that comparisons can be made using up-to-date observations. Indeed, the S08 Supplementary Information recommends this very expedient:

> Since most of the model 20CEN experiments end in 1999, we make the necessary assumption that values of $b_m$ estimated over 1979 to 1999 are representative of the longer-term $b_m$ trends over 1979 to 2006. Examination of the observed data suggests that this assumption is not unreasonable.

However, having described a practical method of testing sensitivity, S08 failed to report the sensitivity of their $d_1^*$ statistic to such updating, a step which we now show yields materially different conclusions than those reported in S08.

## Methods

We downloaded UAH and RSS tropical data for both mid-troposphere (T2/TMT) and lower troposphere (T2LT) levels. We emulated S08 methodology using the statistical language R and in our Supplementary Information we provide turnkey source code and data as used for all calculations.

We benchmarked our emulation of S08 methods against UAH-T2LT data up to 1999, comparing our results to those reported in Tables I and III, yielding a close replication of values. For example, using our emulation, we obtained a $d_1^*$ value of 1.13 for data up to 1999, as compared to the reported value of 1.11 in S08 Table III. Comparisons of S08 to our corresponding emulations for $d_1^*$ statistics are in Table 1 below (see Supplementary Information Table 1 for other S08 Table I and III results). Differences between S08 results and our emulations were slightly greater for the RSS data sets than the UAH data sets. These differences may result from revisions to the 1979-1999 data between the time S08 retrieved them and the time we did. Or they may result from slight differences between our implementation of the methods described in S08 and their actual code. One of the reasons for our archiving both data and source code is to avoid precisely these sorts of pointless reconciliation problems. However, given the very close replication of UAH T2LT results to reported S08 results, we are confident that we have replicated S08 methodology to sufficient accuracy that any slight residual methodological inconsistencies are immaterial to the results reported herein.

**Table 1**
**Comparison of $d_1^*$ Results, 1979-1999**

|          | Reported | Emulated |
|----------|----------|----------|
| UAH T2LT | 1.11     | 1.13     |
| RSS T2LT | 0.37     | 0.49     |
| UAH T2   | 1.19     | 1.36     |
| RSS T2   | 0.44     | 0.63     |

After benchmarking our method, we calculated the $d_1^*$ statistic on a year-by-year basis extending the sample end date from 1999 to 2008 (using December 2008 data, the most recent data available at the time of submission). The results are in Figure 1 below, together with 90th, 95th, and 97.5th percentiles in a *t*-distribution with degrees of freedom calculated according to S08 equation 13.

As S08 reported, their $d_1^*$ statistic was insignificant in all four cases at the 97.5th percentile in 1999. However, that is not the case for more up-to-date data. The $d_1^*$ statistic for the UAH T2 series has exceeded the 97.5th percentile in all years since 2003 (and the 95th percentile since 2001). The UAH T2LT series exceeded the 97.5th percentile in 2008 (and the 95th percentile since 2006.) The $d_1^*$ statistic for the RSS T2 series now exceeds the 90th percentile and appears likely to break the 95th percentile within a few years. Only the $d_1^*$ statistic for RSS T2LT remains insignificant, though it has also increased since 1999.

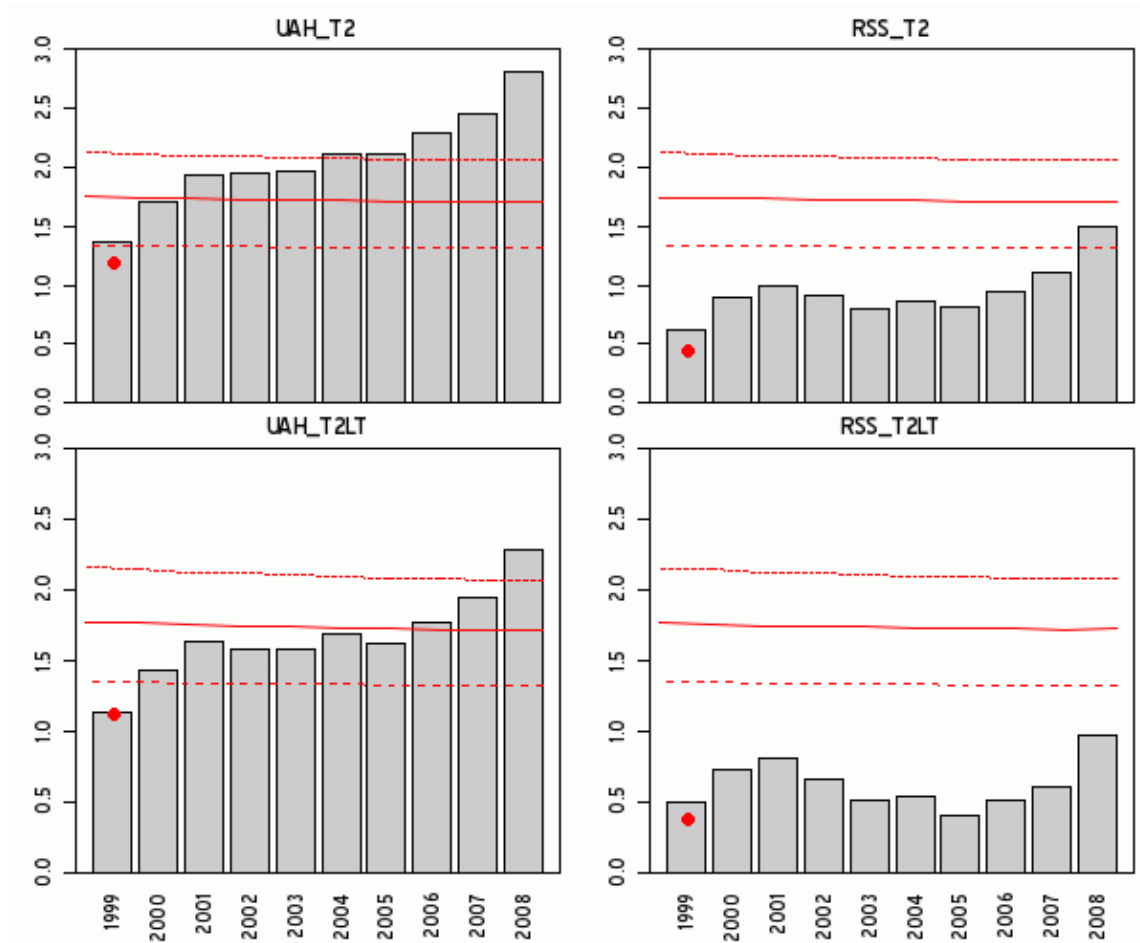

**Figure 1**. $d_1^*$ statistics calculated according to the methods of Santer et al (2008) for UAH/RSS T2 and T2LT series. Red solid – 95th percentile; dotted – 97.5th percentile; dashed – 90th percentile. Red * denotes results reported in Santer et al (2008).

We also examined the impact of S08 autocorrelation adjustments for the question of whether the observed trend is significantly different from zero (see Figure 2 below). Had S08 reported the results of this test on their 1979-1999 dataset, they would have been obliged to note that there was no statistically significant difference between observed trends and zero trend in any of the four cases. But as in the comparisons between observations and model ensemble mean trends, the $d_1^*$ statistic has changed since 1999, though the pattern of change is noticeably different than that shown in Figure 1. Using S08 methodology, there continues to be no statistically significant warming trend in UAH T2 or TLT data. Using the two-sided t-test preferred by S08 (97.5th percentile), the RSS T2 trend still does not have a statistically significant difference from zero trend and is only significant at present using the two-sided t-test that S08 did not use. However, there is a statistically significant difference between the RSS T2LT trend and zero trend under both two-sided and one-sided t-tests.

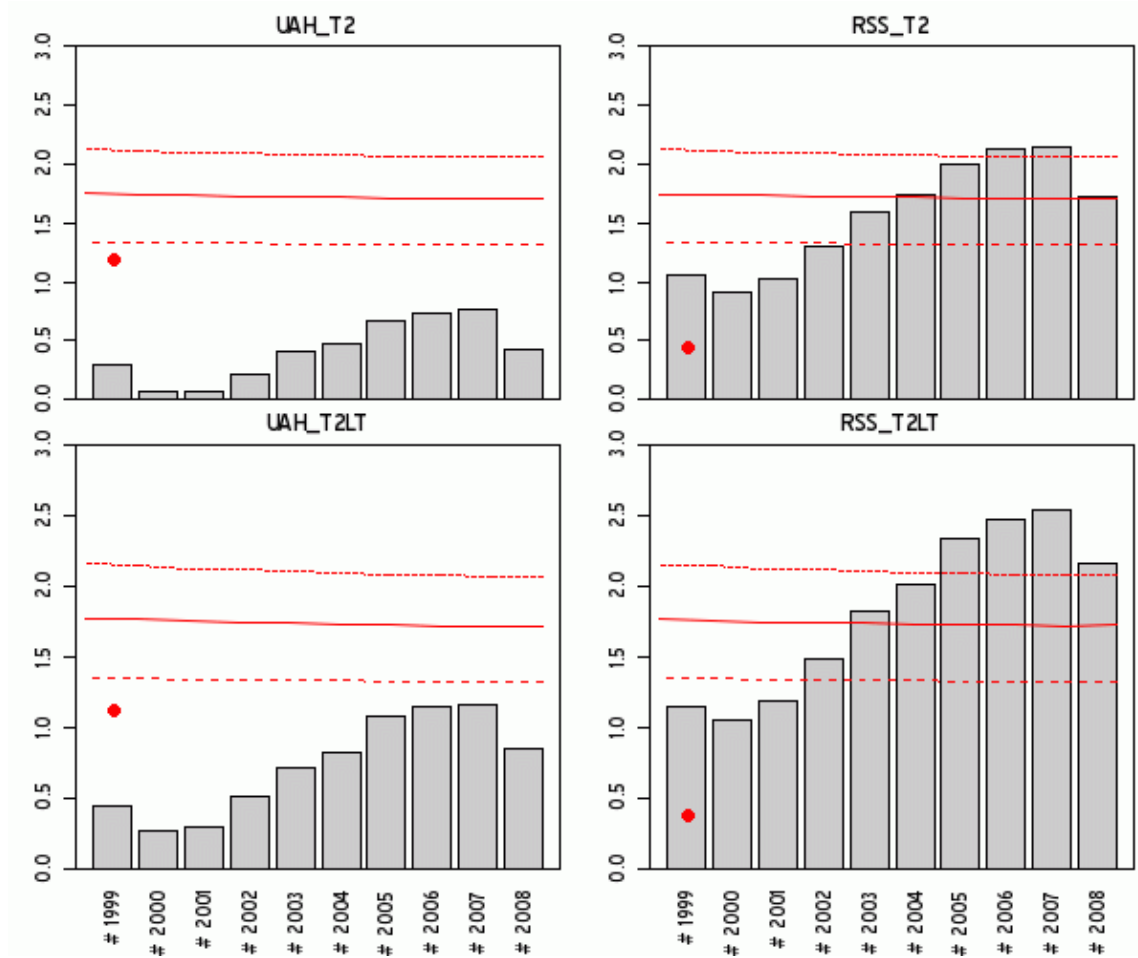

Figure 2. As Figure 1 for null hypothesis of zero trend.

## Discussion

S08 claims to have achieved a "partial resolution" of the long-standing dispute over tropical troposphere temperature trends were premature. Contrary to S08 claims, using their own methodology on up-to-date data, there is a statistically significant difference between model ensemble mean trends and UAH T2 and T2LT observations, with RSS T2 trends verging on a statistically significant difference.

While we agree with S08's criticism of Douglass et al's failure to treat observational trends as random and autocorrelated, ironically, for UAH data, S08 methodology confirms the statistically significant difference between observed trends and model ensemble mean trend that Douglass et al had claimed to show.

It is puzzling that S08 failed to report on the impact of using up-to-date data on their $d_1^*$ statistic measuring the statistical significance of the difference between model ensemble mean trends and observed trends, given the explicit discussion in their Supplementary Information of procedures that readily yield such results.

Our results show one more time that the difference between UAH and RSS trends needs to be resolved. To the extent that this question may have been set aside due to premature S08 claims of a "partial resolution" of this dispute, we urge renewed efforts to assess the relative merits of the UAH and RSS observational data, noting that there are statistical as well as scientific issues involved in their respective interpretations.

All S08 calculations assume that an AR1 autocorrelation model is appropriate for determining the "effective degrees of freedom"; however, nowhere do they prove that this model applies (or provide references to such proofs). Cohn and Lins (2005), Koutsoyiannis (2003) and others have argued that other specifications, in particular models of long term persistence, are appropriate for temperature data. This could affect the estimation of trend standard errors.

After observing the failure of S08 results pertaining to the difference between model ensemble mean trend and observations (their "H2" hypothesis) on up-to-date data, we became interested in testing other S08 findings concerning differences between the population of individual model runs and observations (their "H1" hypothesis).

In Santer et al (2005), S08 coauthors had calculated monthly synthetic data sets for the tropical T2LT and T2 levels for 49 runs (19 models.) S08 stated that they used the same data sets. We requested this data from S08 lead author Santer, who categorically refused to provide it (see http://www.climateaudit.org/?p=4314.) Instead of supplying what would be at most 1 MB or so of monthly data collated by specialists as part of their research work, Santer directed us to the terabytes of archived PCMDI data and challenged us to reproduce their series from scratch. Apart from the pointless and potentially large time cost imposed by this refusal, the task of aggregating PCMDI data with which we are unfamiliar would create the risk of introducing irrelevant collation errors or mismatched averaging steps, leading to superfluous controversy should our results not replicate theirs.

Following this refusal by lead author Santer, we filed a Freedom of Information (FOI) Request to NOAA, applying to coauthors Karl, Free, Solomon and Lanzante. In response, all four denied having copies of any of the model time series used in Santer et al. (2008) and denied having copies of any email correspondence concerning these time series with any other coauthors of Santer et al. (2008). Two other coauthors stated by email that they did not have copies of the series. An FOI request to the U.S. Department of Energy is under consideration.

We asked the editor of this journal to require Santer et al to provide the data, but were advised that the publisher (Royal Meteorological Society) does not presently have a policy requiring data disclosure. We urge the adoption of a modern and effective policy to avoid such disputes in the future.

## REFERENCES:

Cohn, T. A., and H. F. Lins. 2005. Nature's style: Naturally trendy. *Geophys. Res. Lett* 32: 23.

Douglass, D. H., J. R. Christy, B. D. Pearson, and S. F. Singer. 2007. A comparison of tropical temperature trends with model predictions. *Intl J Climatology (Royal Meteorol Soc). DOI* 10.

International Panel on Climate Change. 2007. *Climate Change 2007: The Physical Basis. Review Comments and Responses*. Review Comments and Responses.

Karl, T. R., Susan J. Hassol, Christopher D. Miller, and Willieam L. Murray. 2006. *Temperature Trends in the Lower Atmosphere: Steps for Understanding and Reconciling Differences*. Synthesis and Assessment Product. Climate Change Science Program and the Subcommittee on Global Change Research. http://www.climatescience.gov/Library/sap/sap1-1/finalreport/sap1-1-final-all.pdf.

Koutsoyiannis, D. 2003. Climate change, the Hurst phenomenon, and hydrological statistics/Changement climatique, phénomène de Hurst et statistiques hydrologiques. *Hydrological Sciences Journal/Journal des Sciences Hydrologiques* 48, no. 1: 3-24.

Nychka, D., R Buchberger, T. M. L. Wigley, B. D. Santer, K. E. Taylor, and R.H. Jones. 2000. Nychka, D., Buchberger, R., Wigley, T.M.L., Santer, B.D., Taylor, K.E., Jones, R.H., 2000. Confidence intervals for trend estimates with autocorrelated observations (unpublished manuscript). http://citeseerx.ist.psu.edu/viewdoc/download?doi=10.1.1.33.6828&rep=rep1&type=pdf .

Quenouille, M. H. 1952. *Associated Measurements*. Butterworths Scientific Publications.

Santer, B. D., J. S. Boyle, J. J. Hnilo, K. E. Taylor, T. M. L. Wigley, D. Nychka, D. J. Gaffen, and D. E. Parker. 2000. Statistical significance of trends and trend differences in layer-average atmospheric temperature time series. *Journal of Geophysical Research* 105, no. D6.

Santer, B. D., P. W. Thorne, L. Haimberger, K. E. Taylor, T. M. L. Wigley, J. R. Lanzante, S. Solomon, M. Free, P. J. Gleckler, and P. D. Jones. 2008. Consistency of Modelled and Observed Temperature Trends in the Tropical Troposphere. *Int. J. Climatol*.

Santer, B. D., T. M. L. Wigley, C. Mears, F. J. Wentz, S. A. Klein, D. J. Seidel, K. E. Taylor, P. W. Thorne, M. F. Wehner, and P. J. Gleckler. 2005. *Amplification of*

*Surface Temperature Trends and Variability in the Tropical Atmosphere*. Vol. 309. American Association for the Advancement of Science.

# SUPPLEMENTARY INFORMATION TABLE 1
## Comparison of S08 Results (Tables I and III) to Emulation

|           | Trend      | SE    | SD    | r1    | neff | $d_1^*$ | $d_1$ |
|-----------|------------|-------|-------|-------|------|---------|-------|
|           | (deg C/dec)|       |       |       |      |         |       |
| **Santer**    |        |       |       |       |      |         |       |
| UAH_T2LT  | 0.060      | 0.138 | 0.299 | 0.891 | 14.5 | 1.11    | 7.16  |
| RSS_T2LT  | 0.166      | 0.132 | 0.312 | 0.884 | 15.6 | 0.37    | 2.25  |
| UAH_T2    | 0.043      | 0.129 | 0.306 | 0.873 | 17.1 | 1.19    | 6.78  |
| RSS_T2    | 0.142      | 0.129 | 0.319 | 0.871 | 17.3 | 0.44    | 2.48  |
| **Emulation** |        |       |       |       |      |         |       |
| UAH_T2LT  | 0.059      | 0.136 | 0.300 | 0.888 | 15.0 | 1.13    | 7.19  |
| RSS_T2LT  | 0.150      | 0.131 | 0.308 | 0.882 | 15.8 | 0.49    | 3.01  |
| UAH_T2    | 0.038      | 0.128 | 0.306 | 0.871 | 17.4 | 1.36    | 8.15  |
| RSS_T2    | 0.134      | 0.127 | 0.318 | 0.868 | 17.7 | 0.63    | 3.73  |

# SUPPLEMENTARY INFORMATION
# SOURCE CODE

#Available in ASCII format at http://www.climateaudit.org/scripts/santer

### LOAD DATA
#1. Santer 2008 Table 1

```
    loc="http://www.climateaudit.org/data/models/santer_2008_table1.dat"
    santer=read.table(loc,skip=1)
    names(santer)=c("item","layer","trend","se","sd","r1","neff")
    row.names(santer)=paste(santer[,1],santer[,2],sep="_")
    santer=santer[,3:ncol(santer)]
 #insert Table III info
    santer$d1star=NA;santer$dstar=NA
    santer["UAH_T2LT",c("d1star","dstar")]= c(1.11,7.16) #from Table III for UAH
    santer["RSS_T2LT",c("d1star","dstar")]= c(0.37,2.25) #from Table III for UAH
    santer["UAH_T2",c("d1star","dstar")]= c(1.19,6.78) #from Table III for UAH
    santer["RSS_T2",c("d1star","dstar")]= c(0.44,2.48) #from Table III for UAH
    santer

#                     trend  se    sd    r1    neff  d1star dstar
#HadCRUT3v_TL+O        0.119 0.117 0.197 0.934 8.6   NA     NA
#Multi-model_mean_TL+O 0.146 0.214 0.274 0.915 11.7  NA     NA
#...

####################
## REPLICATE SANTER TABLE 1 FOR MSU
    source("http://www.climateaudit.org/scripts/spaghetti/msu.glb.txt")
    options(digits=3)

    x=msu[,"Trpcs"];id="UAH_T2LT" ; temp=(time(x)>=1979)&(time(x)<2000);
    start0=1979
    x=ts(x[temp],start=c(start0,1),freq=12)
    year=c(time(x))/10;
    N=length(x);N #252
    year=year-mean(year)
    ssx= sum(year^2);ssx#92.6

 #standard deviations
    c(sd(x),santer[id,"sd"]) # 0.300 0.299

 #trend
    fm= lm (x~year)
    c(fm$coef[2],santer[id,"trend"]) # 0.0591 0.0600
```

```
#first principles: SE in summary(fm) is equal to sum(resid^2)/df/ssx
    c(summary(fm)$coef[2,"Std. Error"],sqrt(sum(fm$residuals^2)/fm$df /ssx)  )
        #  0.031 0.031

    (ci=fm$coef[2]+c(-1,1)*qt(.975, fm$df)*summary(fm)$coef[2,"Std. Error"])
        # -0.00193  0.12014

#AR1 coefficient
    r= arima (fm$residuals,order=c(1,0,0))$coef[1]; r # 0.888
    c(r,santer[id,"r1"]) # 0.888 0.891

#neff given AR1
    neff= N * (1-r)/(1+r) ;neff #15
    r1= santer[id,"r1"];r1 #.891
    c(neff,N * (1-r1)/(1+r1) , santer[id,"neff"]) ; #15.0 14.5 14.5

#neff with Nychka formula
    neff_nychka=N*(1-r-.68/sqrt(N))/(1+r+.68/sqrt(N));neff_nychka  #9.08

#se.obs
    c(sqrt(sum(fm$residuals^2)/(santer[id,"neff"]-2)  /ssx), santer[id,"se"] )
       #  0.139 0.138
        #per Santer equation 5

#first principles expression for Santer formula yields very slight differences
    c(santer[id,"se"],
        summary(fm)$coef[2,"Std. Error"] * sqrt( (N-2)/(santer[id,"neff"]-2)) ,
        summary(fm)$coef[2,"Std. Error"] * sqrt( (N-2)/(neff-2)),
        summary(fm)$coef[2,"Std. Error"] * sqrt( (N-2)/(neff_nychka-2))   )
      #  0.138 0.139 0.136 0.184

#first principles expression for t-statistic
    c(summary(fm)$coef[2,"t value"] ,fm$coef[2]/summary(fm)$coef[2,"Std. Error"])
        # 1.91  1.91

#first principles expression for t-statistic with neff
    c( fm$coef[2]/ (summary(fm)$coef[2,"Std. Error"] * sqrt( (N-2)/(neff-2))  ),
       fm$coef[2]/ (summary(fm)$coef[2,"Std. Error"] * sqrt( (N-2)/(neff_nychka-2))  ) )
        # 0.435  0.321
```

#TABLE III CALCULATIONS

```r
#collect model trends and SDs from Santer et al information
   (trend.model=trend.model.T2LT=santer["Multi-model_mean_T2LT",1]); # 0.215
   (sd.model=sd.model.T2LT=santer["Inter-model_S.D._T2LT",1]);   #0.0920

#collect observation trend and SDs as used in Santer et al 2008
   (trend.obs=santer["UAH_T2LT","trend"]) # 0.06
   (se.obs=santer["UAH_T2LT","se"]) #0.138
   M=19  #from Santer
   (neff=santer["UAH_T2LT","neff"])#14.5

#dstar
   c( (trend.model-trend.obs)/ (sd.model/sqrt(M-1)),  santer["UAH_T2LT","dstar"])
    ## 7.15 7.16  #replicates Table III

#d1star
   c( (trend.model-trend.obs)/ sqrt( (sd.model/sqrt(M-1))^2 + se.obs^2),  santer["UAH_T2LT","d1star"])
     #1.11 1.11

#EQUATION 13 #df calculation from Santer equation 13  made into function
   df.d1star=function(sd.model,M,se.obs,neff) {
     C= (sd.model/sqrt(M-1))^2 + se.obs^2;C #  0.0195
     D=(sd.model/sqrt(M-1))^2 /(M-1) + se.obs^2/(neff-2)   ;D # 0.00155
     df.d1star= C/D  #  12.6
     df.d1star}

   df.d1star( sd.model=santer["Inter-model_S.D._T2LT",1],M=19,se.obs=santer["UAH_T2LT","se"], neff=santer["UAH_T2LT","neff"])
       #12.6

   santer$adj_df=NA
   santer["UAH_T2LT","adj_df"]=df.d1star( sd.model.T2LT,M=19,se.obs=santer["UAH_T2LT","se"], neff=santer["UAH_T2LT","neff"])
   santer["RSS_T2LT","adj_df"]=df.d1star( sd.model.T2LT,M=19,se.obs=santer["RSS_T2LT","se"], neff=santer["RSS_T2LT","neff"])
   (trend.model.T2=santer["Multi-model_mean_T2",1]); # 0.199
   (sd.model.T2=santer["Inter-model_S.D._T2",1]);   #0.098
   santer["UAH_T2","adj_df"]=df.d1star( sd.model.T2,M=19,se.obs=santer["UAH_T2","se"], neff=santer["UAH_T2","neff"])
   santer["RSS_T2","adj_df"]=df.d1star( sd.model.T2,M=19,se.obs=santer["RSS_T2","se"], neff=santer["RSS_T2","neff"])
   santer[,"adj_df"]
         #[1]  NA   NA   NA   NA   NA   NA   NA   NA 12.6 13.7   NA   NA 15.2 15.4   NA   NA
```

```
            qt(.95,df=santer["UAH_T2LT","adj_df"])    #    1.78

###MAKE TEHE ABOVE CALCULATIONS INTO A FUNCTION
  #this is a utility function to calculate the items in Santer Table 1 for a given time series
  #this is for monthly data
  #requires x - monthly observations; id - Santer target

  f=function(x,end0=2050,start0=1979,M=19,
        trend.model=trend.model.T2LT, sd.model=sd.model.T2LT ) {
     f=rep(NA,9);
names(f)=c("trend","se","sd","r1","neff","dstar","d1star","adj_df","d1zero")
        temp= (time(x)< (end0+1))&(time(x)>=start0)
        x=ts(x[temp],start=c(start0,1),freq=12)
        year=c(time(x))/10;  N=length(x); year=year-mean(year)
        f["sd"]=sd(x)
        fm= lm (x~year)
        trend.obs=f["trend"]= fm$coef[2]
        r=f["r1"]= arima (fm$residuals,order=c(1,0,0))$coef[1]; #
        neff=f["neff"]= N * (1-r)/(1+r) ;
        se.obs=f["se"] = sqrt((N-2)/(neff-2))* summary(fm)$coef[2,"Std. Error"]
        f["d1star"]= (trend.model-trend.obs)/ sqrt( (sd.model/sqrt(M-1))^2 + se.obs^2)
        f["dstar"]= (trend.model-trend.obs)/ (sd.model/sqrt(M-1))
        f["adj_df"]= df.d1star(sd.model,M,se.obs,neff)
        f["d1zero"]=f["trend"]/f["se"]
        f
        }

#######
        santertest=rep(list(NA),4);names(santertest)=c("UAH_T2LT","RSS_T2LT","UAH_T2","RSS_T2")
        emulation=rep(list(NA),4);names(emulation)=c("UAH_T2LT","RSS_T2LT","UAH_T2","RSS_T2")

#########

############
##A. MSU_T2LT
###############

##1999
        source("http://www.climateaudit.org/scripts/spaghetti/msu.glb.txt")
        index0=c("trend","se", "sd",    "r1", "neff", "d1star","dstar","adj_df")
```

```
        emulation[["UAH_T2LT"]]=rbind(santer["UAH_T2LT",index0],f(msu[,"Trpcs"],
end0=1999)[index0]);
        row.names(emulation[["UAH_T2LT"]])=c("santer","emulation");emulation[["UA
H_T2LT"]]

        #          trend   se    sd    r1  neff d1star dstar adj_df
        #santer    0.0600 0.138 0.299 0.891 14.5  1.11  7.16  12.6
        #emulation 0.0591 0.136 0.300 0.888 15.0  1.13  7.19  13.1

##NOW FOR UPDATE
        tropics=msu[,"Trpcs"]
        santertest[["UAH_T2LT"]]= santer["UAH_T2LT",index0]
        santertest[["UAH_T2LT"]]$d1zero=NA
        index=c(index0,"d1zero")
        for (i in 1999:2008) {
         santertest[["UAH_T2LT"]]=
rbind( santertest[["UAH_T2LT"]],f(msu[,"Trpcs"],end0= i)[index] )
         }
        row.names(santertest[["UAH_T2LT"]])=paste("#",c("santer",1999:2008),sep="")
        santertest[["UAH_T2LT"]]

#        trend   se     sd    r1  neff d1star dstar adj_df d1zero
#santer 0.0600 0.1380 0.299 0.891 14.5  1.11  7.16  12.6   NA
#1999   0.0591 0.1359 0.300 0.888 15.0  1.13  7.19  13.1  0.435
#2000   0.0328 0.1255 0.297 0.887 15.7  1.43  8.40  13.8  0.262
#2001   0.0326 0.1101 0.292 0.878 17.9  1.63  8.41  15.9  0.296
#2002   0.0516 0.1016 0.291 0.878 18.7  1.57  7.54  16.8  0.507
#2003   0.0657 0.0925 0.290 0.873 20.3  1.57  6.88  18.3  0.711
#2004   0.0687 0.0845 0.288 0.869 21.9  1.68  6.75  19.7  0.813
#2005   0.0839 0.0785 0.289 0.867 23.2  1.61  6.05  20.9  1.069
#2006   0.0824 0.0718 0.287 0.861 25.0  1.77  6.12  22.5  1.147
#2007   0.0781 0.0673 0.285 0.861 26.1  1.94  6.31  23.3  1.161
#2008   0.0557 0.0665 0.286 0.866 25.9  2.28  7.34  23.2  0.839

##S08: This test is two-tailed, since we have no expectation a priori
#regarding the direction of the trend difference. #after equation 5

    #two-sided 2 test at 5% p-level
       qt(c(.975),df=santertest[["UAH_T2LT"]]$adj_df)
             # [1] 2.17 2.16 2.15 2.12 2.11 2.10 2.09 2.08 2.07 2.07 2.07

    #one-sided 2 test at 5% p-level
```

```
            qt(c(.95),df=santertest[["UAH_T2LT"]]$adj_df)
            #  1.78 1.77 1.76 1.75 1.74 1.73 1.73 1.72 1.72 1.71 1.71
```

############
## B. UAH T2
###################

```
        source("http://www.climateaudit.org/scripts/spaghetti/msu.t2.txt")

 ###EMULATION
        tropics=ts.union(tropics, msu.t2[,"Trpcs"])
        emulation[["UAH_T2"]]=rbind(santer["UAH_T2",index0],f(msu.t2[,"Trpcs"],end0=1999)[index0]);
        row.names(emulation[["UAH_T2"]])=c("santer","emulation");emulation[["UAH_T2"]]

#          trend   se   sd   r1 neff d1star dstar adj_df
#santer    0.0430 0.129 0.306 0.873 17.1  1.19  6.78  15.2
#emulation 0.0383 0.128 0.306 0.871 17.4  1.36  8.15  15.4

        santertest[["UAH_T2"]]= santer["UAH_T2",index0]
        santertest[["UAH_T2"]]$d1zero=NA
        for (i in 1999:2008) {
         santertest[["UAH_T2"]]= rbind( santertest[["UAH_T2"]],f(msu.t2[,"Trpcs"],end0= i)[index] )
        }
        row.names(santertest[["UAH_T2"]])=paste("#",c("santer",1999:2008),sep="")
        santertest[["UAH_T2"]]

#        trend   se    sd   r1 neff d1star dstar adj_df d1zero
#santer 0.04300 0.1290 0.306 0.873 17.1  1.19  6.78  15.2    NA
#1999   0.03830 0.1279 0.306 0.871 17.4  1.36  8.15  15.4 0.2995
#2000   0.00889 0.1189 0.303 0.872 18.0  1.70  9.51  16.0 0.0747
#2001   0.00736 0.1055 0.299 0.864 20.1  1.93  9.58  18.1 0.0698
#2002   0.02155 0.0965 0.295 0.862 21.3  1.96  8.92  19.2 0.2233
#2003   0.03638 0.0884 0.292 0.859 22.8  1.96  8.24  20.6 0.4114
#2004   0.03838 0.0810 0.289 0.856 24.3  2.11  8.15  21.9 0.4740
#2005   0.05019 0.0748 0.288 0.852 25.8  2.12  7.60  23.2 0.6711
#2006   0.05054 0.0686 0.285 0.847 27.8  2.29  7.58  24.8 0.7371
#2007   0.04910 0.0640 0.282 0.845 29.1  2.46  7.65  25.8 0.7675
#2008   0.02739 0.0631 0.285 0.851 29.0  2.81  8.65  25.7 0.4343
```

############
## C. RSS TLT

```
###################
        source("http://www.climateaudit.org/scripts/spaghetti/tlt3.glb.txt")
        tropics=ts.union(tropics,tlt3[,"20.20"])

  ###EMULATION
        emulation[["RSS_T2LT"]]=rbind(santer["RSS_T2LT",index0],f(tlt3[,"20.20"],end0=1999)[index0]);
        row.names(emulation[["RSS_T2LT"]])=c("santer","emulation");emulation[["RSS_T2LT"]]

#          trend   se   sd   r1 neff d1star dstar adj_df
#santer    0.166 0.132 0.312 0.884 15.6  0.370  2.25   13.7
#emulation 0.150 0.131 0.308 0.882 15.8  0.493  3.01   13.9

        santertest[["RSS_T2LT"]]= santer["RSS_T2LT",index0]
        santertest[["RSS_T2LT"]]$d1zero=NA
        for (i in 1999:2008) {
         santertest[["RSS_T2LT"]]=
rbind( santertest[["RSS_T2LT"]],f(tlt3[,"20.20"],end0= i)[index] )
          }
        row.names(santertest[["RSS_T2LT"]])=paste("#",c("santer",1999:2008),sep="")
        santertest[["RSS_T2LT"]]

#        trend   se    sd   r1 neff d1star dstar adj_df d1zero
#santer  0.166 0.1320 0.312 0.884 15.6  0.370  2.25   13.7    NA
#1999    0.150 0.1306 0.308 0.882 15.8  0.493  3.01   13.9   1.15
#2000    0.126 0.1204 0.303 0.882 16.5  0.727  4.10   14.6   1.05
#2001    0.127 0.1071 0.300 0.876 18.3  0.802  4.04   16.3   1.19
#2002    0.148 0.0999 0.303 0.877 18.9  0.651  3.07   16.9   1.49
#2003    0.167 0.0917 0.308 0.874 20.2  0.506  2.20   18.2   1.82
#2004    0.169 0.0841 0.307 0.871 21.5  0.529  2.12   19.4   2.01
#2005    0.182 0.0780 0.311 0.869 22.7  0.406  1.51   20.5   2.33
#2006    0.177 0.0716 0.309 0.865 24.4  0.508  1.75   21.9   2.47
#2007    0.172 0.0678 0.308 0.866 25.0  0.606  1.99   22.4   2.54
#2008    0.146 0.0677 0.306 0.872 24.6  0.973  3.19   22.1   2.15

############
##D. RSS TMT
###################
        source("http://www.climateaudit.org/scripts/spaghetti/rss.tmt.txt")
        tropics=ts.union(tropics,tmt[,"20.20"])
        tropics=cbind(time(tropics),tropics)
```

```
        dimnames(tropics)[[2]]=c("year","UAH_T2LT","UAH_T2","RSS_T2LT","RSS_T2")
        ##write.table(tropics,file="d:/climate/data/models/santer.tropics.dat",sep="\t",row.names=FALSE)

  ###EMULATION
        emulation[["RSS_T2"]]=rbind(santer["RSS_T2",index0],f(tmt[,"20.20"],end0=1999)[index0]);
        row.names(emulation[["RSS_T2"]])=c("santer","emulation");emulation[["RSS_T2"]]
#          trend   se    sd    r1   neff d1star dstar adj_df
#santer    0.142 0.129 0.319 0.871 17.3  0.440  2.48   15.4
#emulation 0.134 0.127 0.318 0.868 17.7  0.626  3.73   15.8

        santertest[["RSS_T2"]]= santer["RSS_T2",index0]
        santertest[["RSS_T2"]]$d1zero=NA
        for (i in 1999:2008) {
         santertest[["RSS_T2"]]= rbind( santertest[["RSS_T2"]],f(tmt[,"20.20"],end0=i)[index] )
         }
        row.names(santertest[["RSS_T2"]])=paste("#",c("santer",1999:2008),sep="")
        santertest[["RSS_T2"]]

#       trend   se    sd    r1   neff d1star dstar adj_df d1zero
#santer 0.142 0.1290 0.319 0.871 17.3  0.440  2.48   15.4   NA
#1999   0.134 0.1273 0.318 0.868 17.7  0.626  3.73   15.8   1.05
#2000   0.107 0.1179 0.312 0.869 18.4  0.899  4.97   16.5   0.91
#2001   0.108 0.1050 0.308 0.862 20.4  1.000  4.95   18.4   1.03
#2002   0.125 0.0967 0.309 0.861 21.5  0.907  4.15   19.4   1.29
#2003   0.142 0.0891 0.311 0.859 22.8  0.795  3.36   20.6   1.60
#2004   0.142 0.0816 0.308 0.856 24.2  0.861  3.35   21.9   1.74
#2005   0.151 0.0752 0.309 0.853 25.8  0.823  2.97   23.2   2.00
#2006   0.146 0.0689 0.307 0.847 27.7  0.949  3.16   24.8   2.12
#2007   0.139 0.0649 0.304 0.848 28.7  1.104  3.49   25.4   2.15
#2008   0.112 0.0651 0.305 0.855 28.1  1.493  4.73   25.0   1.73

##########
##TABLE 3  RESTATED
##########

   #Reported Table III
      k=1;sapply(santertest[c(2,1,4,3)], function(A) A[k,c("dstar","d1star")])
      #      RSS_T2LT UAH_T2LT RSS_T2 UAH_T2
      #dstar  2.25     7.16    2.48   6.78
      #d1star 0.37     1.11    0.44   1.19
```

```
#1999 Emulation gives very close match for UAH T2LT; higher values for other data
	k=2;sapply(santertest[c(2,1,4,3)], function(A) A[k,c("dstar","d1star")])
	#		RSS_T2LT UAH_T2LT RSS_T2 UAH_T2
	#dstar	3.01	7.19	3.73	8.15
	#d1star	0.493	1.13	0.626	1.36

#To end 2007
	k=4;sapply(santertest[c(2,1,4,3)], function(A) A["#2007",c("dstar","d1star")])
	#	RSS_T2LT UAH_T2LT RSS_T2 UAH_T2
	#dstar	1.99	6.31	3.49	7.65
	#d1star	0.606	1.94	1.10	2.46

#To end Nov 2008
	sapply(santertest[c(2,1,4,3)], function(A) A["#2008",c("dstar","d1star")])
	#	RSS_T2LT UAH_T2LT RSS_T2 UAH_T2
	#dstar	3.14	7.32	4.66	8.62
	#d1star	0.954	2.26	1.47	2.79

#d1star collation
	d1star=sapply(santertest[c(1,3,2,4)], function(A) A[2:11,"d1star"])
	row.names(d1star)=paste("#",1999:2008);d1star
	row.names(d1star)=1999:2008

#	UAH_T2LT UAH_T2 RSS_T2LT RSS_T2
# 1999	1.13	1.36	0.493	0.626
# 2000	1.43	1.70	0.727	0.899
# 2001	1.63	1.93	0.802	1.000
# 2002	1.57	1.96	0.651	0.907
# 2003	1.57	1.96	0.506	0.795
# 2004	1.68	2.11	0.529	0.861
# 2005	1.61	2.12	0.406	0.823
# 2006	1.77	2.29	0.508	0.949
# 2007	1.94	2.46	0.606	1.104
# 2008	2.28	2.81	0.973	1.493

#########
##d1star PERCENTILES RESTATED
###################
	for (i in 1:4)
santertest[[i]]$pt_d1star=pt(santertest[[i]]$d1star,santertest[[i]]$adj_df)
	test=sapply(santertest[c(2,1,4,3)], function(A) A[,"pt_d1star"])
	row.names(test)=paste("#",c("santer",1999:2008));test
```

```
#       RSS_T2LT UAH_T2LT RSS_T2 UAH_T2
# santer  0.641   0.856   0.667  0.874
# 1999    0.685   0.861   0.730  0.904
# 2000    0.761   0.913   0.809  0.946
# 2001    0.783   0.938   0.835  0.965
# 2002    0.738   0.933   0.812  0.967
# 2003    0.691   0.933   0.782  0.968
# 2004    0.699   0.945   0.801  0.977
# 2005    0.655   0.939   0.791  0.977
# 2006    0.692   0.955   0.824  0.985
# 2007    0.725   0.967   0.860  0.989
# 2008    0.829   0.984   0.926  0.995

#GDD(file=file.path("d:/climate/images/2009","santer.fig1.gif"), type="gif", w=600, h=500)
  nf=layout(array(1:4,dim=c(2,2)),heights=c(1,1.4) )
  par0=list(c(0,3,2,1), c(4,3,2,1), c(0,3,2,1),c(4,3,2,1))
  index=c("UAH_T2", "UAH_T2LT","RSS_T2","RSS_T2LT")
  for (k in 1:4 ) {
       par(mar=par0[[k]])
       if (!(k%%2 == 0))
barplot(d1star[,index[k]],col="grey80",las=3,font=2,ylim=c(0,3),ylab="",names.arg="")
else barplot(d1star[,index[k]],col="grey80",las=3,font=2,ylim=c(0,3),ylab="");
       box()
       #mtext(side=2,font=2,"Adjusted t-stat",line=2)
       target=qt(.95,df=santertest[[index[k]]]$adj_df[2:11])
       lines(seq(0,12,12/9),target,col=2)
       target=qt(.9,df=santertest[[index[k]]]$adj_df[2:11])
       lines(seq(0,12,12/9),target,col=2,lty=2)
       target=qt(.975,df=santertest[[index[k]]]$adj_df[2:11])
       lines(seq(0,12,12/9),target,col=2,lty=3)
       title(paste(index[k]))
       #legend("topleft",lty=c(3,1,2),col=2,legend=c(.975,.95,.9),cex=.6)
       points(.75,santertest[[index[k]]]$d1star[1],pch=19,col=2,cex=1.4)
       }
#dev.off()

###Zero
     d1zero=sapply(santertest[c(3,1,4,2)], function(A) A[2:11,"d1zero"])
     row.names(d1zero)=1999:2008
```

```
#       UAH_T2 UAH_T2LT RSS_T2 RSS_T2LT
# 1999 0.2995   0.435   1.05   1.15
# 2000 0.0747   0.262   0.91   1.05
# 2001 0.0698   0.296   1.03   1.19
# 2002 0.2233   0.507   1.29   1.49
# 2003 0.4114   0.711   1.60   1.82
# 2004 0.4740   0.813   1.74   2.01
# 2005 0.6711   1.069   2.00   2.33
# 2006 0.7371   1.147   2.12   2.47
# 2007 0.7675   1.161   2.15   2.54
# 2008 0.4343   0.839   1.73   2.15

#GDD(file=file.path("d:/climate/images/2009","santer.fig2.gif"), type="gif", w=600, h=500)
  nf=layout(array(1:4,dim=c(2,2)),heights=c(1,1.4) )
  par0=list(c(0,3,2,1), c(4,3,2,1), c(0,3,2,1),c(4,3,2,1))
  index=c("UAH_T2", "UAH_T2LT","RSS_T2","RSS_T2LT")
  for (k in 1:4 ) {
        par(mar=par0[[k]])
        if (!(k%%2 == 0))
barplot(d1zero[,index[k]],col="grey80",las=3,font=2,ylim=c(0,3),ylab="",names.arg="")
else barplot(d1zero[,index[k]],col="grey80",las=3,font=2,ylim=c(0,3),ylab="");
        box()
        #mtext(side=2,font=2,"Adjusted t-stat",line=2)
        target=qt(.95,df=santertest[[index[k]]]$adj_df[2:11])
        lines(seq(0,12,12/9),target,col=2)
        target=qt(.9,df=santertest[[index[k]]]$adj_df[2:11])
        lines(seq(0,12,12/9),target,col=2,lty=2)
        target=qt(.975,df=santertest[[index[k]]]$adj_df[2:11])
        lines(seq(0,12,12/9),target,col=2,lty=3)
        title(paste(index[k]))
        #legend("topleft",lty=c(3,1,2),col=2,legend=c(.975,.95,.9),cex=.6)
        points(.75,santertest[[index[k]]]$d1star[1],pch=19,col=2,cex=1.4)
        }
#dev.off()
```